\documentclass[fleqn,usenatbib]{mnras}

\usepackage[T1]{fontenc}

\DeclareRobustCommand{\VAN}[3]{#2}
\let\VANthebibliography\thebibliography
\def\thebibliography{\DeclareRobustCommand{\VAN}[3]{##3}\VANthebibliography}

\usepackage{graphicx}	
\usepackage{amsmath}	
\usepackage{amssymb}	

\title[The stellar populations of backsplash galaxies]
{Exploring the stellar populations of backsplash galaxies}

\author[Ferreras et al.]
{I. Ferreras$^{1,2,3}$\thanks{E-mail: iferreras@iac.es}, 
A. B\"ohm$^{4}$, K. Umetsu$^{5}$, V. Sampaio$^{6,7}$, R.~R. de Carvalho$^{6}$\\
$^1$ Instituto de Astrof{\'i}sica de Canarias, Calle V{\'i}a L{\'a}ctea s/n,
E38205, La Laguna, Tenerife, Spain\\
$^2$ Department of Physics and Astronomy, University College London, London WC1E 6BT, UK\\
$^3$ Departamento de Astrof{\'i}sica, Universidad de La Laguna, E38206 La Laguna, Tenerife, Spain\\
$^4$ Department of Astrophysics, University of Vienna, 1180, Vienna, Austria\\
$^5$ Academia Sinica Institute of Astronomy and Astrophysics (ASIAA), No. 1, Section 4, Roosevelt Road, 10617, Taipei, Taiwan\\
$^6$ NAT-Universidade Cidade de S\~{a}o Paulo, Rua Galv\~{a}o Bueno, 868, 01506-000 S\~{a}o Paulo, SP, Brazil\\
$^7$ School of Physics and Astronomy, University of Nottingham, University Park,
Nottingham NG7 2RD, UK\\
}

\date{Accepted 2022 December 21. Received 2022 December 12; in original form 2022 July 4}

\pubyear{2023}

\begin{document}
\label{firstpage}
\pagerange{\pageref{firstpage}--\pageref{lastpage}}
\maketitle

\begin{abstract}
  Backsplash galaxies are those that traverse and overshoot cluster
  cores as they fall into these structures. They are affected by
  environment, and should stand out in contrast to the infalling
  population. We target galaxies in the vicinity of
  clusters (R$\gtrsim$R$_{\rm 200}$) and select a sample in projected
  phase space (PPS), from the compilation of Sampaio et al. based on
  SDSS data.  We present a statistical analysis, comparing two regions
  in PPS, with the same projected distance to the cluster but
  different velocity. The analysis relies on the presence of
  variations in the stellar population content of 
  backsplash galaxies. We find a lower limit in the fractional contribution of 
  $\sim$5\% with respect to the general sample
  of infalling galaxies at similar group-centric distance when using
  single line strength analysis, or $\sim$15-30\% when adopting
  bivariate distributions. The stellar populations show a subtle but
  significant difference towards older ages, and a higher fraction of
  quiescent galaxies. We also compare this set with a general field
  sample, where a substantially larger difference in galaxy properties
  is found, with the field sample being consistently younger, metal
  poorer and with a lower fraction of quiescent galaxies. Noting that
  our ``cluster'' sample is located outside of the virial radius, we
  expect this difference to be caused by pre-processing of the
  infalling galaxies in the overall higher density regions.
\end{abstract}



\section{Introduction}
\label{Sec:Intro}

The evolution of cosmic structure is mostly
driven by the growth of dark matter fluctuations in an
expanding Universe, that takes shape as an entangled
web, where fluctuations cover a wide range of scales.
In this picture, galaxies are located within dark matter
halos, where gas cools down and forms stars. While the
overall properties of galaxies strongly correlate with ``local''
observables  -- such as morphology, stellar mass or central velocity
dispersion -- the mass distribution over
larger scales (i.e. ``the environment'') also  play
an important role: In high density regions, such as
the cores of clusters, it is more likely to find 
galaxies with an 
early-type morphology, and with overall older stellar
ages with respect to the field, suggesting that
the process of galaxy formation is ``accelerated'' in
denser places \citep[e.g.][]{Dressler:97}.
A number of environment-related mechanisms
have been proposed to explain the observations in groups
and clusters, including ram pressure stripping, strangulation,
harassment. Those processes affect the evolution of the gas
reservoir in galaxies and their subsequent star formation
\cite[see, e.g.,][and references therein]{AP:15}. Disentangling
the different effects is one of the main goals in extragalactic
astrophysics, as they link the evolution of galaxies with the
larger dark matter halos that host
them \citep[see, e.g.,][]{Weinmann:06,Peng:10, Henriques:17}.

Stellar populations can be used as a tracer of these processes by
comparing their properties in carefully selected samples that would
appear homogeneous if the environment did not affect their star
formation histories.  Many papers present detailed analyses of
spectroscopic indicators of stellar age and
metallicity \citep[to name a few:][]{Thomas:05,Rogers:10,AP:10,FLB:11,FG:15,Goddard:17,Scholz:22},
with the major conclusion being that the largest difference in the
stellar population content concerns the central vs satellite nature of
these galaxies, with the former having slightly younger ages,
representative of extended episodes of star formation, in contrast
with satellites, that feature older populations, indicative of
efficient quenching. Beyond the central/satellite dichotomy, the
effect of other environment properties -- such as host halo mass, or
cluster-centric distance -- is
discernible \citep[see, e.g,][]{Smith:06,Smith:12,Taranu:14,AP:19,Vitor:21}. 
However, the overall effect on the underlying stellar populations
appears weaker when the samples
are segregated by morphology or evolutionary phase
\citep[e.g.,][]{FLB:14,Trussler:21}.
In early-type galaxies, environment
effects are subdominant to the more important scaling relations
with local observables such as stellar mass or velocity dispersion
\citep[e.g.][]{SAMIGrad:19}.
Some targeted differences can be found, for instance, in close galaxy
pairs, where satellites of the same mass, orbiting more massive
primaries tend to be slightly older, a potential indicator of
assembly bias \citep{CPs:17,CPs:19}. A more detailed view in
projected phase space finds subtle but consistent differences
between the stellar populations of galaxies in clusters with a
Gaussian or non Gaussian velocity distribution \citep{Ribeiro:13,RRdC:17},
reflecting the state of virialisation, and thus the intensity of
environment-related effects on the star formation histories
\citep{Vitor:21}.

Backsplash galaxies are defined as those that
already fell into the cluster, passing through the dense cluster
core, and overshot the structure, so that they are found at larger
distances, typically outside of the virial radius \citep{Gill:05}.
In this framework, backsplash galaxies constitute very interesting
laboratories, as they represent systems that have been strongly 
affected by environment-related effects, but, due to
their motion within the structure, 
are found in the outer, lower density regions, where such effects
are not expected \citep{Balogh:00}. Therefore, the population of backsplash
galaxies would stand out 
with respect to the other galaxies in the same region, mainly
infallers. Moreover, galaxies falling towards a cluster were likely
part of a lower mass group, and therefore were also affected by the 
environment of the earlier structures, defined as 
pre-processsing \citep[e.g.,][]{Fujita:04}.
Numerical simulations show that a high fraction of galaxies
outside of the virial radius have already plunged within the core of the
corresponding cluster, thus being affected by their environment
\citep{Mamon:04}, losing a substantial fraction of their mass
\citep{Gill:05}. Observational studies based on cluster samples
from SDSS confirm the presence of backsplash galaxies at
projected radii $\sim$1--2\,R$_{\rm 200}$ \citep{Pimbblet:11},
with signs of quenched star formation activity \citep{Mahajan:11,Muriel:14}.

The presence of backsplash galaxies is related to a more fundamental
property concerning the formation of dark matter
halos \citep[e.g.][]{FG:84}. Due to the collisionless nature of dark matter, 
during the collapse and virialisation of a structure, the particles will
overshoot the nascent core and will reach some distance before falling
back again. Such behaviour explains the steep density gradients
found in the outer profile of dark matter halos \citep[e.g.][]{DK:14,Ad:14,Xhakaj:20},
which somehow represents a limit of the bound particles with respect to
the infalling matter. This signature can also be found in velocity
space \citep{Okumura:18}. Similarly to dark matter,  
galaxies that fall into
a cluster but overshoot them, reach an orbital apocenter further out, 
defining the so-called splashback radius (R$_{sp}$) that
could be detected in the radial profile of the galaxy distribution,
where a sharp drop is expected outside of this radius.
\citet{More:15} estimate  
R$_{\rm sp}$ between 0.8 and 1.5\,R$_{200m}$ (with R$_{200m}$ defined at 
200 times the mean density of the Universe), depending on the accretion
history of the cluster, and therefore allowing for the determination
of assembly bias, since R$_{sp}$ will depend on halo parameters
other than its mass (see also \citealt{BW:17,Sunayama:19}).
Recently, a number of attempts have been made to determine the
location of R$_{\rm sp}$ in galaxy clusters from weak lensing and galaxy
clustering observations
\citep[e.g.,][]{More:16,Baxter:17,Umetsu:17,Chang:18,Contigiani:19,Shin:19,Murata:20}

Comparisons with numerical simulations allow us to characterize in
detail the properties of the backsplash population and its connection
to assembly bias \citep{BW:17}, finding that over half of all galaxies
located, at present, within R$_{200}$ and 2\,R$_{200}$ of the cluster
centre have already passed through the core -- where R$_{200}$ is
defined as the radius where the density reaches 200 times the critical
value -- with this population assembled over relatively recent times,
therefore being dependent on the recent formation history of the
cluster \citep{Haggar:20}.

This paper focuses on a methodology to contrast the spectroscopic
signatures of stellar populations in a sample selected in projected
phase space, in order to explore the effect of environment in
backsplash galaxies, and constrain their fractional contribution to
the samples located just outside of the virial radius. Our methodology
relies on variations of the stellar population content, therefore
assumes that the effect of the passage through the cluster leaves some
imprint on the absorption line spectra. Section~\ref{Sec:PPS} defines the two key subsets adopted to
find backsplash galaxies. Section~\ref{Sec:StPops} explores a number
of observable properties, focusing on three targeted line strengths
that are strongly dependent on the stellar age of the stellar
populations. Section~\ref{Sec:Cumul} describes a simple methodology
based on cumulative distributions to assess the fraction of backsplash
galaxies.  Finally, Section~\ref{Sec:Conc} summarises our main
conclusions.

\section{Defining the backsplash region in projected phase space}
\label{Sec:PPS}

Our sample of backsplash candidates is retrieved from the compilation
of \citet{Vitor:21}, who made use of the 
galaxy group catalogue of \citet{Yang:07}, based on the 
 Sloan Digital Sky Survey \citep[][hereafter SDSS]{York:00}.
This catalogue classifies galaxies in projected phase space (PPS)
relative to their respective groups, where the
peculiar, line-of-sight, velocity of a galaxy -- measured as a
fraction of the velocity dispersion within the group --  is plotted
against its 2D projected group-centric distance. By making comparisons with
computer simulations, the PPS diagram can be sliced in a number
of zones that characterize the dynamical state of the galaxy within
the group, and can give the time since the galaxy fell in that
structure \citep[see, e.g.][]{Oman:16,Rhee:17,AP:19}. The \citet{Vitor:21} sample
targets \citet{Yang:07} groups with mass M$_{200}>10^{14}$M$_\odot$, therefore we
can refer to these systems as clusters.
The properties of the 319  clusters are estimated via
Virial Analysis \citep{Lopes:09} applied to the final member list
defined by the Shiftgapper technique \citep{RRdC:17}. Note that
only members within R$_{200}$ are considered in the estimate of the
velocity dispersion. In this paper, the radius that characterises
each group is given by R$_{\rm 200}$ defined as the position where
the density is 200 times the critical value, as laid out in
\citet{Lopes:09}.

\begin{figure}
\includegraphics[width=85mm]{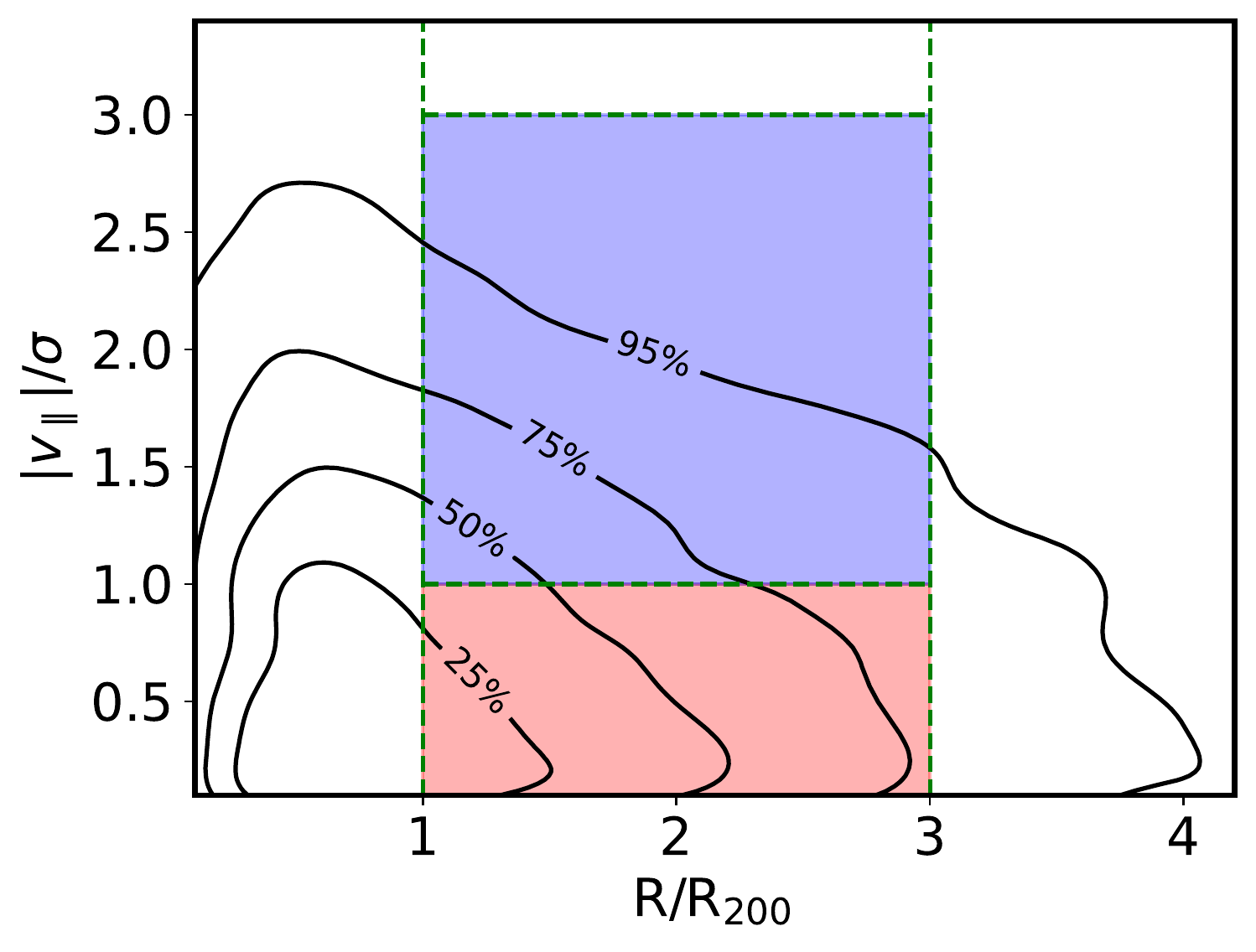}
\caption{Projected phase space diagram showing contours that engulf a
  given fraction of the total sample, as labelled.  The dashed lines
  mark the region defined for this paper to detect backsplash
  galaxies, located outside of the virial radius of the structures,
  split into high (blue) and low (red) line of sight velocity
  (compared with respect to the velocity dispersion of the
  structure). The sample in the blue region is expected to have a
  higher fraction of backsplash galaxies, with respect to the general
  population of field/infalling galaxies. Hereafter, all figures will
  use this colour coding to separate the two subsamples.}
\label{fig:PPS}
\end{figure}

Fig.~\ref{fig:PPS} shows the PPS diagram of the sample, as a density
plot. The original set consists of 570,643 galaxies, from which we
have 23,631 satellites with detailed phase space information.  Modulo
projection effects, backsplash galaxies should be located outside of
the group radius. According to \citet{Haggar:20}, over half of all
galaxies located between one and two group radii (measured as
R$_{200}$), are backsplash galaxies, as measured at z=0 in a set of
detailed numerical simulations. This fraction decreases to negligible
values at R$\gtrsim$3\,R$_{200}$. Moreover, this fraction increases in
dynamically relaxed clusters (i.e. Gaussian clusters). Our criterion
to split the sample on the PPS diagram is aimed at testing whether the
spectroscopic signature of the backsplash galaxies can be determined
by focusing on the not-too-distant outskirts of clusters, split with
respect to relative velocity.  The expected quenching effects will be
noticeable in the signature of the stellar populations, although we
should also expect additional factors affecting the stellar
populations of the infalling galaxies. The sample is also split with
respect to the dynamical state of the group, into Gaussian (G) and non
Gaussian (nG) clusters, a criterion based on the distribution of
velocities. The latter represent structures with a more recent
dynamical formation history and a higher diversity regarding galaxy
properties, suggesting higher levels of pre-processing
\citep[][]{RRdC:17,RP:17,Vitor:21}. We note that Gaussian clusters are
already virialised systems (at least within R$_{\rm 200}$), where the infall of
new members is, in principle, smaller than in non-Gaussian
clusters. Therefore, in the vicinity of nG grous we expect to find a
higher fraction of infalling galaxies, thus with overall higher levels
of pre-processing, as these infallers belong to smaller groups that
are in the process of merging. We also note that the fraction of interlopers
-- defined as galaxies that are within 1-3\,R$_{200}$ in projection, but not
in 3D radial distance, $r_{200}$, can be rather high \citep[see figure~2 of][]{Oman:16}.
However, we will see below that the PPS-selected sample shows strong differences
with respect to the field.

\begin{figure}
\includegraphics[width=85mm]{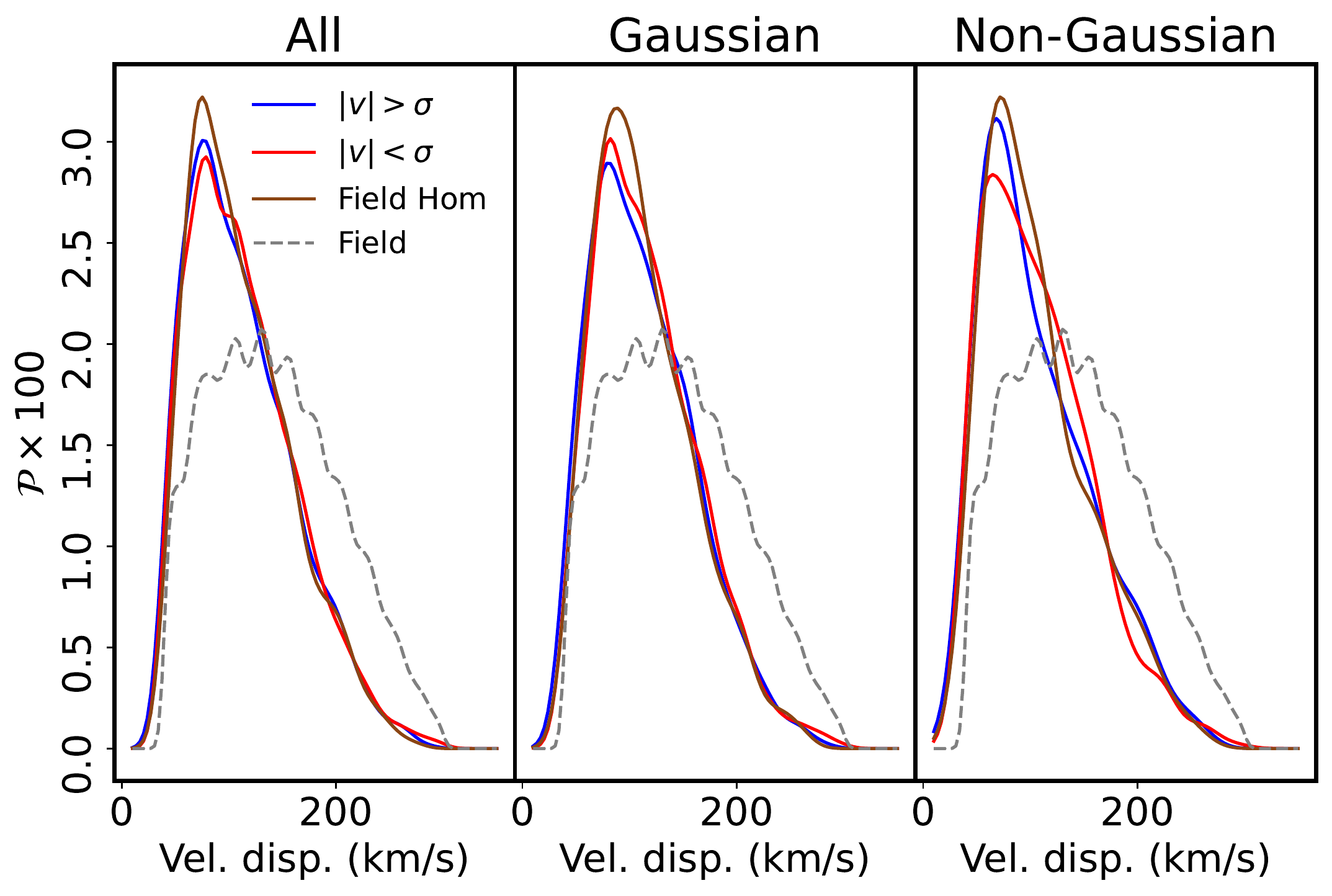}
\caption{Distribution of stellar velocity dispersion in the subsamples
  explored in this paper. The ``field'' reference is also shown in brown,
  including the distribution before (dotted) and after (solid)
  homogeneisation.}
  \label{fig:vdisp}
\end{figure}

We select galaxies in the region of PPS: R/R$_{\rm 200}\in [1,3]$
and $|v_\parallel|/\sigma<3$ (a total of 12,738 SDSS spectra), split into
high- and low-velocity galaxies at $|v_\parallel|=\sigma$, as highlighted in 
blue and red, respectively, in Fig.~\ref{fig:PPS}. We follow this colour
coding throughout the paper. 
From this set, we retrieve the line strength measurements
from the SDSS {\sc galSpecIndx} table, as defined in 
\citet{Jarle:04}. The cross-correlation produces 11,252 galaxies,
with 5,282 (1,951) classified as belonging to Gaussian (non Gaussian)
clusters. The remaining 4,019 cannot be accurately classified regarding
the Gaussian nature of the velocity dispersion of the cluster, but are also
included in a general sample that comprise all cluster galaxies with well-defined PPS
information. As comparison with a general (``field'') sample, we also extract
from SDSS (Legacy) a complete set of spectra that are unconstrained in PPS, imposing
a high enough S/N ($>$5\footnote{Defined by the {\tt snMedian\_r} parameter
in the SpecObjAll table as the median signal-to-noise over all good pixels
in the SDSS $r$ band, given per pixel, $\sim$1\AA.}) in the same redshift window, 0.03$<$z$<$0.10.
Moreover, we exclude from this sample all galaxies that, according to the groups
classification  of \citet{Yang:07}, inhabit a halo with mass above $10^{14}$\,M$_\odot$.
This control sample comprises over 200 thousand spectra and represents the
general population of galaxies not located in high density regions.
We note that projected distances R$<$R$_{\rm 200}$ will include galaxies
outside of the virial sphere, and thus will include infallers and backsplash
galaxies as well. For instance, \citet{Rines:05} estimate 20\% of absorption
line galaxies and 50\% of emission line galaxies to be interlopers inside the
virial cylinder. However, we aim at a robust assessment that minimises all
systematics associated with the sample selection, focusing on projected
cluster-centric distances R>R$_{\rm 200}$, and thus fully eliminating 
galaxies inside the cluster.

\begin{figure*}
  \includegraphics[width=85mm]{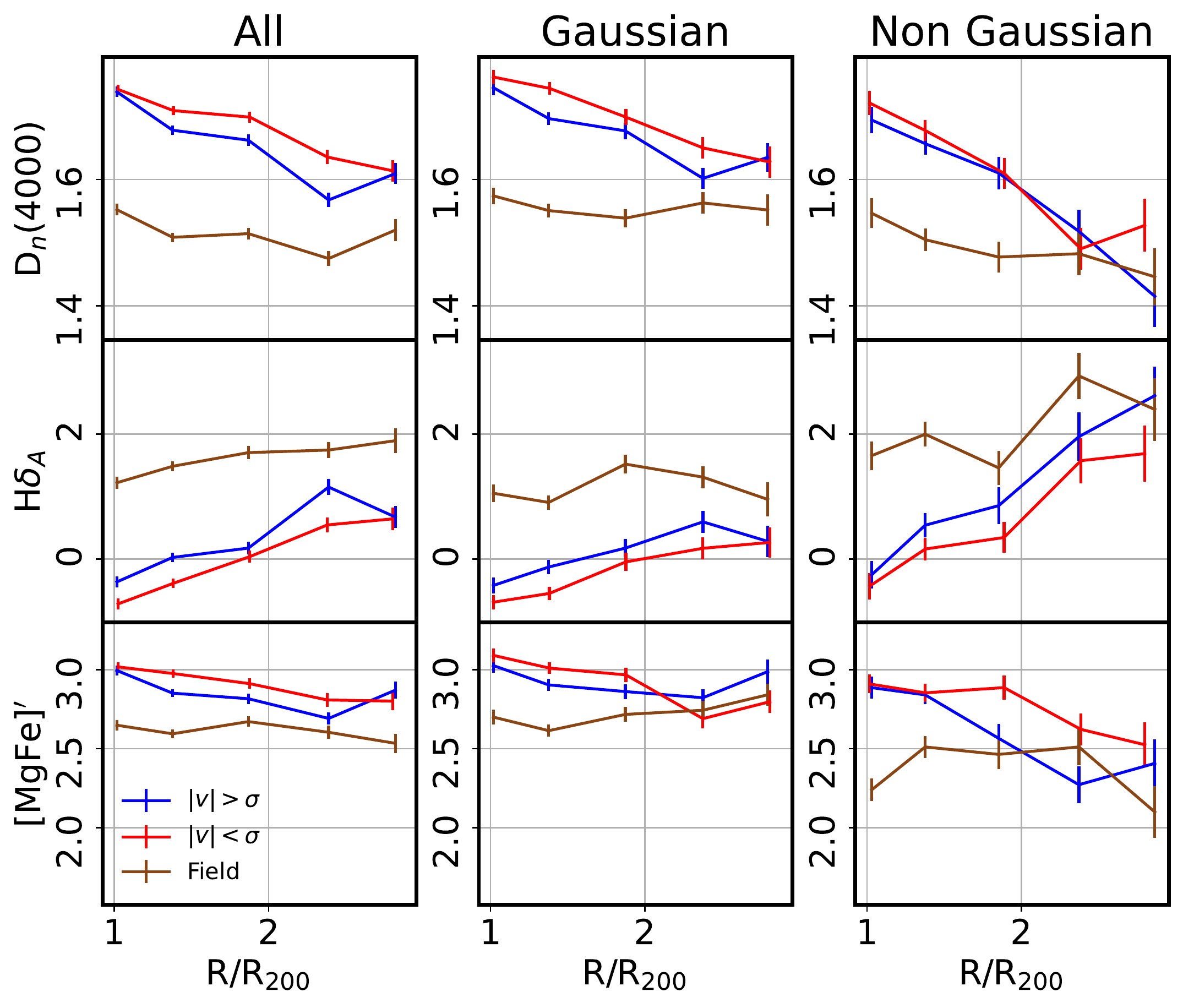}
  \includegraphics[width=85mm]{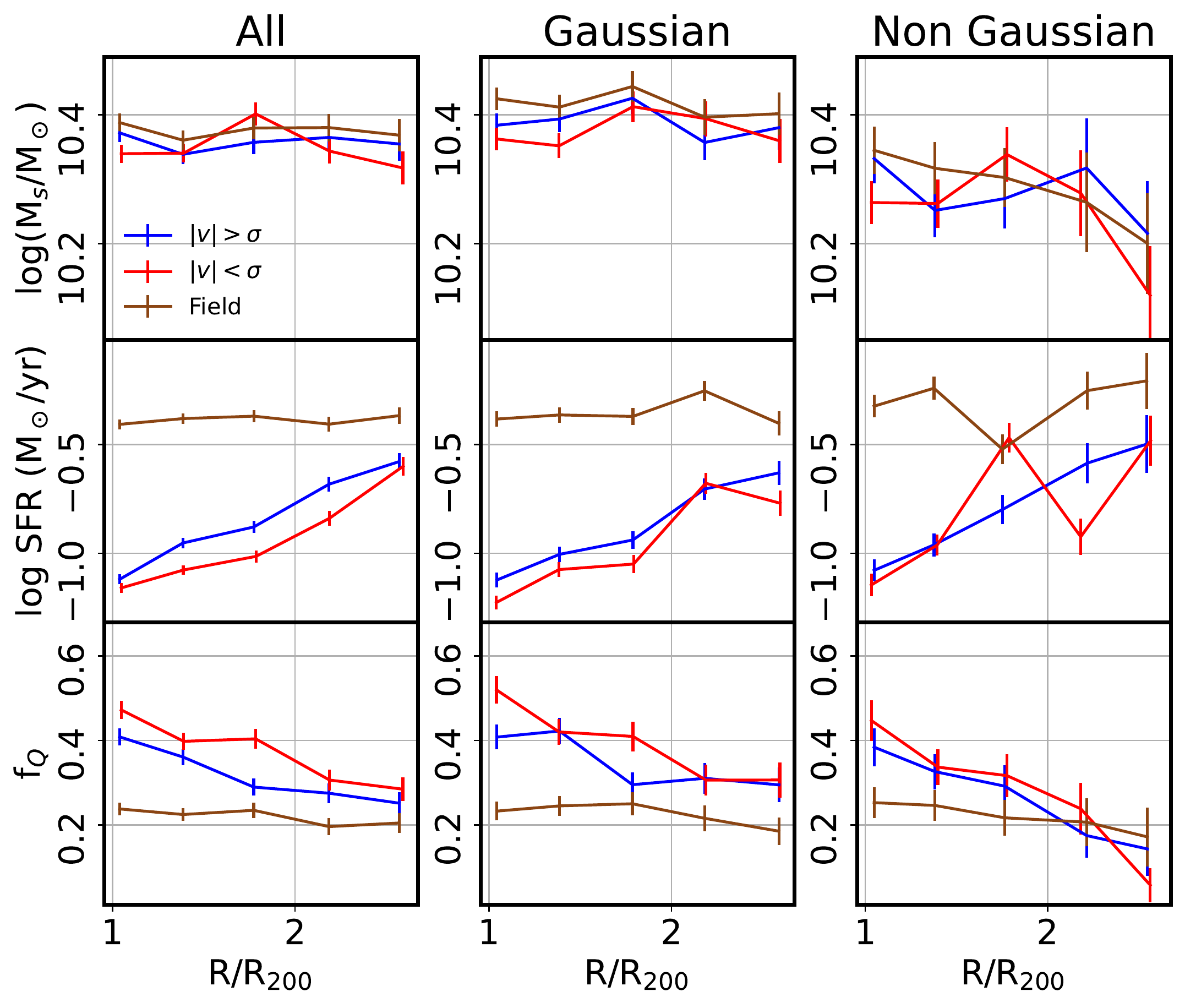}
\caption{Trends with respect to cluster-centric distance, measured as a
  fraction of the group radius (R$_{\rm 200}$). In each panel, the samples with high 
  and low $|v_\parallel/\sigma|$ are shown in blue and red, respectively.
  The brown lines correspond to the field sample of SDSS galaxies selected
  with the same stellar velocity dispersion as the group-related subsamples
  in each radial bin. We emphasize that the homogenisation process is
  done in individual bins, to make sure we remove any trends in the stellar
  populations caused by differences in the distribution of (internal)
  velocity dispersion.
  The datapoints correspond to the median within each bin, and
  the error bars represent the 1\,$\sigma$ error in the median. The leftmost
  (rightmost) figures show spectral line strengths, and a number of general
  galaxy observables, as labelled. In each figure, the sample is shown separately
  (in columns) for the whole sample, and for galaxies in Gaussian and non-Gaussian
  groups (see text for details).}
\label{fig:EWs}
\end{figure*}

Since this analysis focuses on the comparison of line strengths
that constrain the stellar population properties, it is important to ensure that
all samples feature the same distribution of stellar velocity dispersion, which
consistently appears as the dominant driver of population properties in galaxies
\citep[see, e.g.][]{Bernardi:03, SAMIGrad:19}. We extract subsamples from the
original ones, so that the cross-comparison sets involve galaxies with identical 
distributions of velocity dispersion.
This homogeneisation process is illustrated in Fig.~\ref{fig:vdisp},
where we compare the distributions (from left to right), of the complete sample,
galaxies in Gaussian clusters and in non-Gaussian clusters, respectively. In
each panel, galaxies with high- (low-) $|v_\parallel|/\sigma$ are shown in blue (red).
The original general sample from SDSS (i.e. ``field'') is shown as a brown dashed line,
with a clear excess at high velocity dispersion with respect to the
PPS-constrained samples. The distribution of the homogenised field
sample appears as a solid brown line.  We note that hereafter, in each
comparison, all samples are homogenised with respect to the smaller
subset, by random selection of galaxies until the target distribution
is reached.

\section{Probing the stellar population properties}
\label{Sec:StPops}

Fig.~\ref{fig:EWs} (left) shows three of the line strengths -- from top
to bottom: D$_n$(4000), H$\delta_A$ 
and [MgFe]$^\prime$.
The 4000\AA\ break strength follows the definition of the index
as a ratio of flux redward and blueward of the break, as defined in
\citet{Bruzual:83}, but over a narrower spectral window, as
proposed by \citet{Balogh:99}. 
The Balmer line uses the definition of \citet{WO:97}, and 
has been corrected
for emission contamination: we use the  {\tt lick{\_}hd{\_}sub} parameter in
the SDSS {\sc galSpecIndx} table, which removes all emission lines detected
in the spectra at the 3$\sigma$ level \citep{Jarle:04}.
The [MgFe]$^\prime$ index is
defined as ${\rm \sqrt{Mgb (0.72 Fe5270 + 0.28\ Fe5335)}}$ \citep{TMB:03},
where those indices are the standard Lick definitions shown
in \citet{Trager:98}.

The results are
shown in bins regarding group-centric distance.  The samples within
each radial bin are homogeneised in stellar velocity dispersion as
described above. The red and blue lines correspond to subsamples with
a high and low projected velocity, consistently with the shaded zones
of Fig.~\ref{fig:PPS}. For reference, we include as brown lines the
results for a general (i.e. field) sample of SDSS galaxies with the
same distribution in velocity dispersion and stellar mass as the
targeted samples. Within each bin in R/R$_{\rm 200}$, we extract a sample
of SDSS galaxies from the general catalogue, with the same velocity
dispersion as the reference, which is always the smaller sample, i.e.
the high $|v_\parallel|/\sigma$ subset in that radial bin. Therefore,
while the R/R$_{\rm 200}$ has no actual meaning for the SDSS general
sample it simply represents a subset with the same stellar velocity
dispersion as the group-related samples in the same bin.
Variations in the field sample
regarding group-centric distance only reflect differences in the
stellar velocity dispersion of the cluster galaxies. Note
the substantial difference towards higher
D$_n$(4000) and [MgFe]$^\prime$, and lower H$\delta_A$,
in the low $|v_\parallel|/\sigma$ sample, 
corresponding to older and more metal rich populations, suggesting
these as the typical populations in backsplash galaxies.

As reference, Fig.~\ref{fig:EWs} (right) shows the trends of 
stellar mass, star formation rate and fraction of quiescent galaxies
($f_Q$, defined as the fraction of galaxies with a BPT classification flag $-1$ according to the
{\sc galspecExtra} catalogue, along with weak specific star formation, sSFR$<$0.03\,Gyr$^{-1}$).
The samples are indistinguishable regarding
stellar mass. Note that the homogenisation process
(Fig.~\ref{fig:vdisp}) somehow forces this result in general, to ensure that
the differences in stellar population properties are not caused by a
systematic difference in stellar velocity dispersion. The star formation rate does show a
substantial difference with respect to $|v_\parallel|/\sigma$, with the
low velocity sample having consistently weaker star formation activity, a result also
confirmed by the increased fraction of quiescent galaxies ($f_Q$). In all the plots of
Fig.~\ref{fig:EWs} it is also evident the remarkable difference between galaxies
in the ``catchment area'' of groups (i.e. both blue and red lines) and the
general population extracted from SDSS (in brown). The latter shows substantially
weaker D$_n$(4000), [MgFe]$^\prime$, and stronger H$\delta_A$, reflecting
younger and metal-poorer populations, and a correspondingly higher star formation
rate and lower quiescent fraction. We emphasize that, by construction, the general
SDSS sample is defined to have the same distribution of velocity dispersion as
those selected in PPS. Hence, this difference should
be interpreted as pre-processing in the higher density regions surrounding
galaxy groups. Therefore, we see two levels of variation regarding these samples: the
strong difference from pre-processing (brown vs blue/red), along with the subtle difference
introduced by backsplash galaxies (blue vs red).

\begin{figure}
\includegraphics[width=85mm]{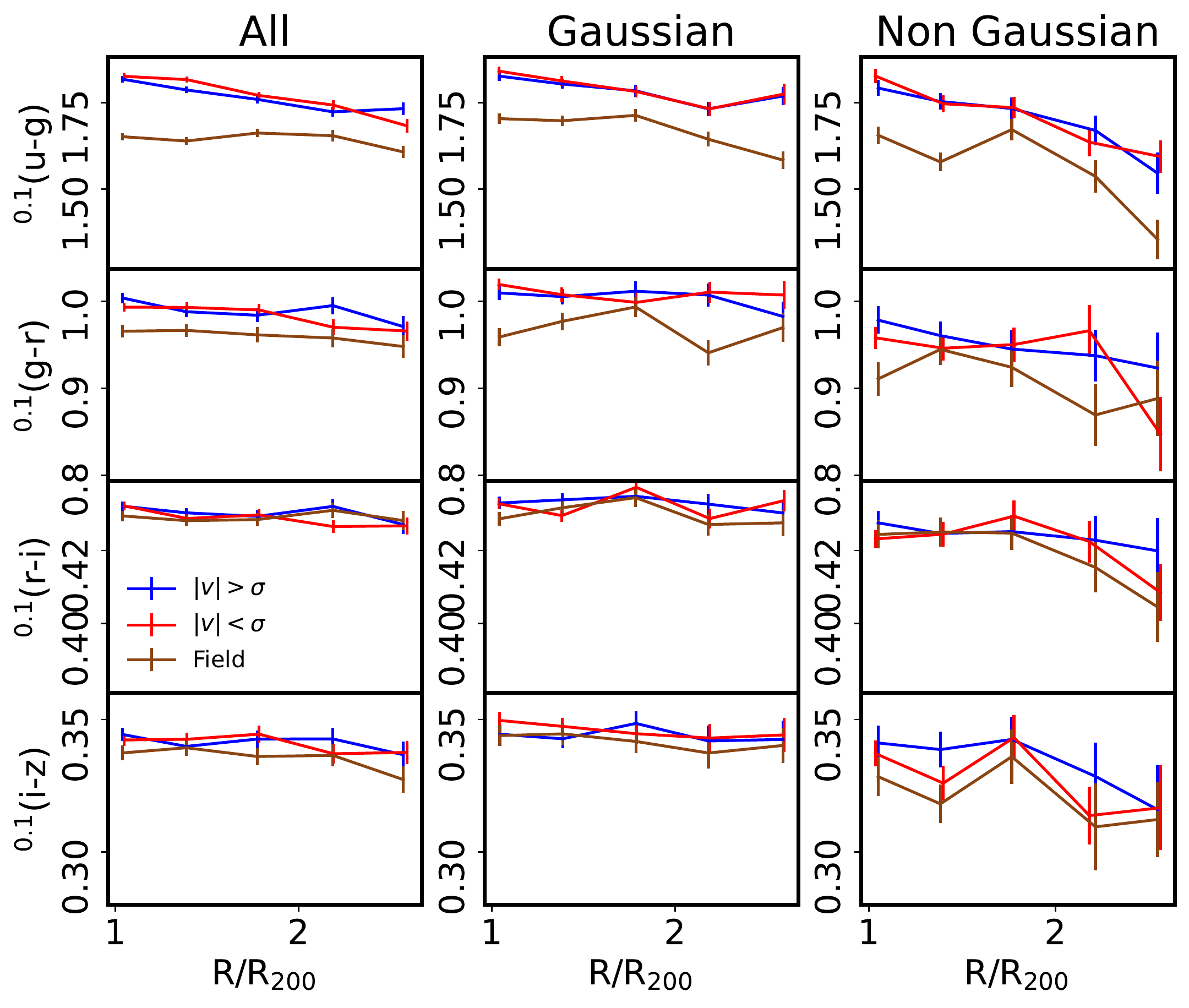}
\caption{Equivalent of Fig.~\ref{fig:EWs}, showing the radial trends of SDSS colours
  with respect to cluster-centric distance. The colours are k-corrected to z=0.1, and
  are measured within the 3\,arcsec fibre diameter of the SDSS classic spectrograph.
}
\label{fig:Clr}
\end{figure}

In addition, Fig.~\ref{fig:EWs} shows the results for the full set of
galaxies selected in PPS, and for galaxies in Gaussian and
non-Gaussian clusters. Note that the number of galaxies in  non-Gaussian clusters
is smaller in our sample, so that the Poisson noise is higher. The
trends are rather similar between G and nG clusters. While simulations suggest
a higher fractional contribution from backsplash galaxies in Gaussian
clusters -- i.e. already virialised systems that had more time to
produce overshooting orbits -- the effect on the line strengths is not
so evident, perhaps reflecting a weaker environment impact on the
stellar populations by the cluster passage, or a more substantial
amount of pre-processing in nG clusters, thus producing a higher
quenched fraction that counterbalances the additional contribution
of backsplash galaxies in G clusters \citep{Vitor:21}. It is worth noting that
the comparison includes the Balmer absorption index H$\delta_A$, which
is especially sensitive to changes in the stellar population content
within the recent $\sim 1$\,Gyr, i.e. comparable with typical
dynamical timescales in these systems.

Fig.~\ref{fig:Clr} shows the trends of the standard colours from the
SDSS photometric measurements -- after correcting for foreground, Milky Way, dust
extinction, taking the fluxes within the 3\,arcsec
diameter fibers, and k-correcting the colours to a common redshift
z=0.1, following \citet{BR:07}. The group vs field difference is once
more evident, with bluer colours in the field sample. However, the
more subtle changes regarding backsplash are undistinguishable with
broadband photometry.

\begin{figure}
\includegraphics[width=85mm]{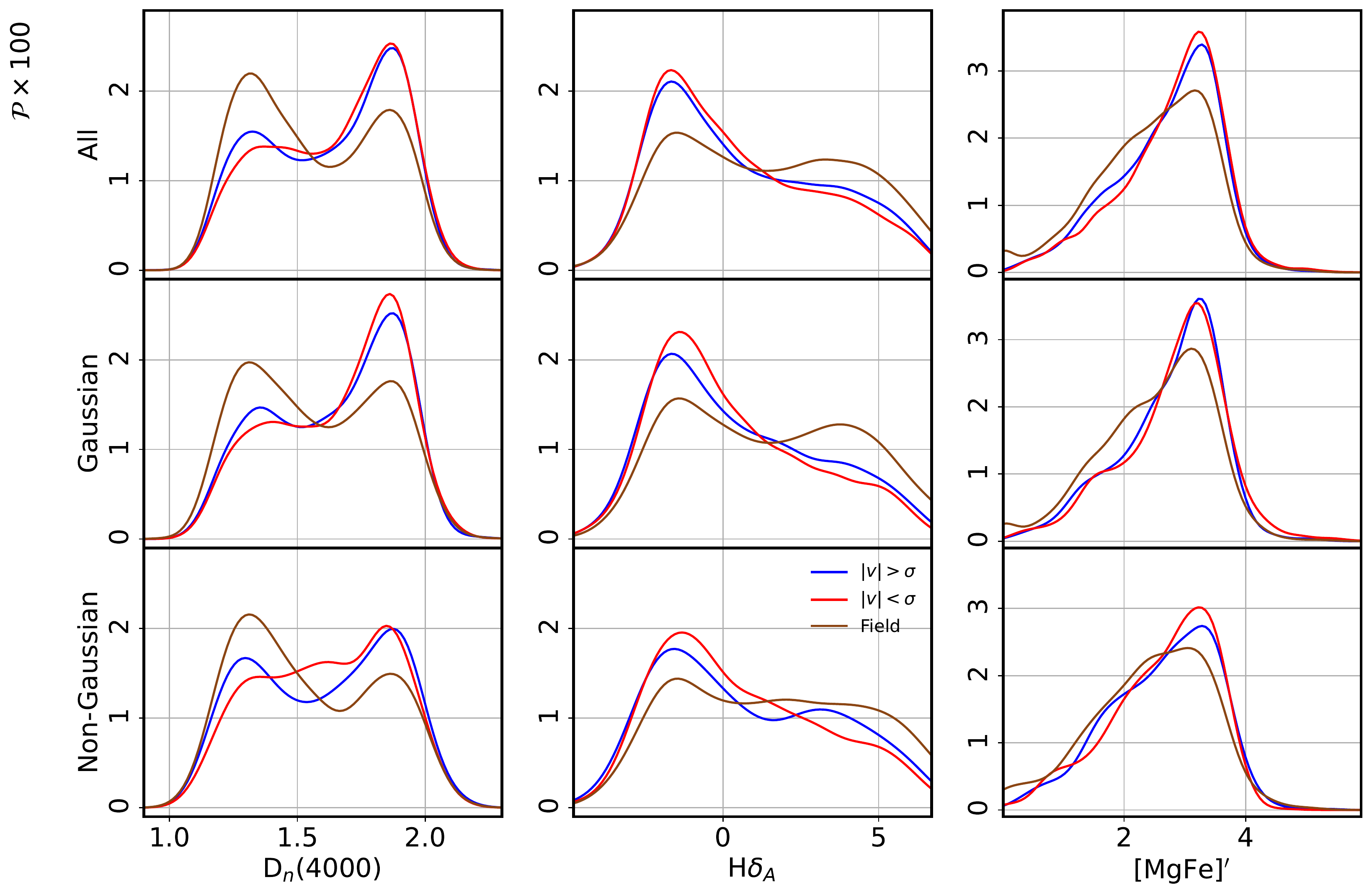}
\caption{Histogram of line strengths in galaxies
  chosen with projected radial distance
  R$\in$[1,3]\,R$_{\rm 200}$, with high- (low-)
  velocity galaxies in blue (red). The field sample
  from SDSS, homogenised in stellar velocity
  dispersion, is shown in brown.}
  \label{fig:hist}
\end{figure}

Complementary to Fig.~\ref{fig:EWs} -- where the median of the distributions
were shown as a function of projected group-centric distance -- we show in 
Fig.~\ref{fig:hist} the distribution of targeted line strengths, taking
all galaxies within the interval R/R$_{\rm 200}\in[1,3]$, following the same
split regarding $|v_\parallel|/\sigma$. The well-known bimodality
\citep[e.g.][]{Strateva:01,Baldry:04,JA:19} is evident
in the 4000\AA\ break strength, with a more prominent blue cloud (i.e. the
peak at low D$_n$(4000) in the field sample. Within groups, the low 
velocity subset has a slightly stronger red sequence, i.e. once more reflecting
an additional component of quenched galaxies. This behaviour is also prominent
in H$\delta_A$. We apply the Anderson-Darling test for k-samples \citep{AD_ksamp}  
to statistically confirm the difference between
the subsets regarding velocity, finding low significance ($p$) values
for the general cluster sample in D$_n$(4000) ($p$=0.065) and H$\delta_A$ ($p<$0.001)
thus rejecting the null hypothesis that both sets originate from the same
parent distribution. Differences regarding the metallicity-sensitive
index [MgFe]$^\prime$ are only substantial between field and groups
(being higher in the latter). However, they are not conclusive when
splitting group galaxies into high- and low-velocity. We will see
below that a bivariate plot involving [MgFe]$^\prime$ and D$_n$(4000)
produces a cleaner separation between these two sets (i.e. extending
the analysis beyond a simple, one-dimensional interpretation, akin to
that of a Simple Stellar Population).

\section{The contribution of backsplash galaxies}
\label{Sec:Cumul}

\begin{figure}
\includegraphics[width=85mm]{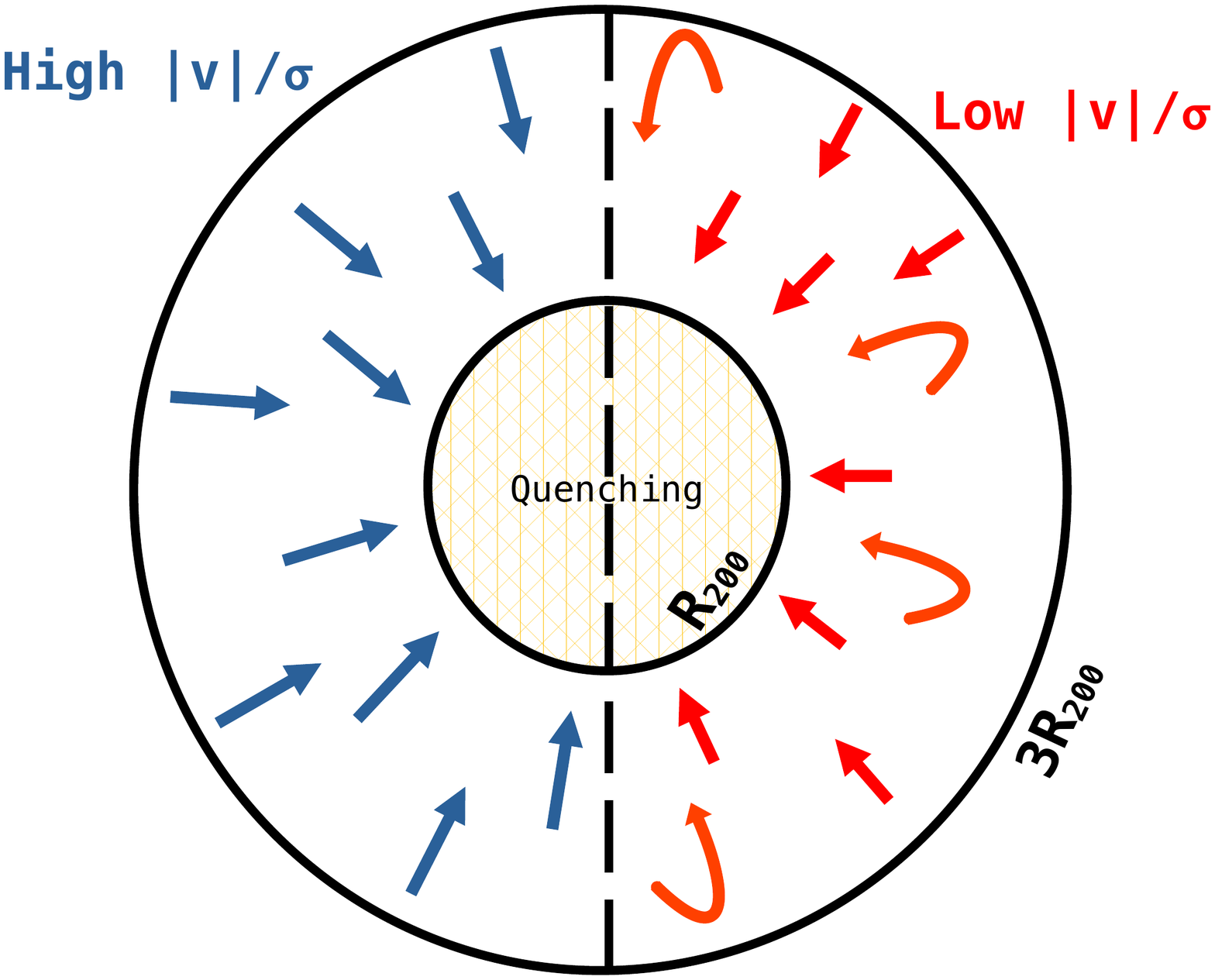}
\caption{Sketch of the ansatz adopted in our analysis
  of backsplash galaxies. We assume that only galaxies with
  low projected velocities ($|v|<\sigma$) include the backsplash
  population, in addition to infalling galaxies. Within the cluster
  (R$<$R$_{200}$) we assume star formation is quenched.}
  \label{fig:model}
\end{figure}

In the outskirts of groups and clusters, we expect to find
galaxies in different ``dynamical phases'' concerning their
motion with respect to those. We simplify this scenario by
assuming that the mixture reduces to two 
components in our working sample of galaxies at R$\gtrsim$R$_{\rm 200}$,
namely: 1) infalling galaxies with a substantial amount of
pre-processing -- relative to a field sample of galaxies with similar
stellar velocity dispersion, and 2) backsplash galaxies that have
undergone strong environment-related effects during (at least one)
passage through the cluster core. When split with respect to projected
velocity, we expect a higher fraction of backsplash galaxies at low
$|v_\parallel|/\sigma$.  Unfortunately, the effect of the cluster
passage on the absorption spectra may be subtle, so that we can not
pick individual galaxies as representative of this sample.  We follow
instead a statistical estimate, imposing the ansatz that
backsplash galaxies can {\sl only} be
found in the $|v_\parallel|<\sigma$ subsample (see Fig.~\ref{fig:model}). We emphasize this is
just an approximation, but it will allow us to produce robust lower
bounds on the contribution of backsplash galaxies in clusters, as
long as the effect of the group/cluster passage leaves discernible
imprints on the spectra.
Previous estimates impose specific models 
about how this population should look like. For instance,
\citet{Pimbblet:11} adopt a mixture model, imposing the backsplash
population to have the same quenched features as those found in
galaxies at the cores of the
clusters, leading to a fraction around 56\% at the virial radius,
slowly decreasing outwards. However, such an approach would imply that
the infalling population should have no quenched star formation, 
to be properly discerned in the mixture,  
whereas pre-processing of the infallers will also produce 
quenched galaxies. In fact, Fig.~\ref{fig:EWs} confirms a large
difference between the field sample and those in the vicinity of
clusters -- regardless of their velocity --
hence showing that the presence of quenched features is
prevalent even {\sl before} the galaxy enters the cluster.  Therefore,
we emphasize that if we choose the field sample as reference,
any method based on a prior on quenching signatures
is expected to overestimate the fraction of backsplash galaxies, as
they will be contaminated by (quenched) infallers.

\begin{figure}
\includegraphics[width=85mm]{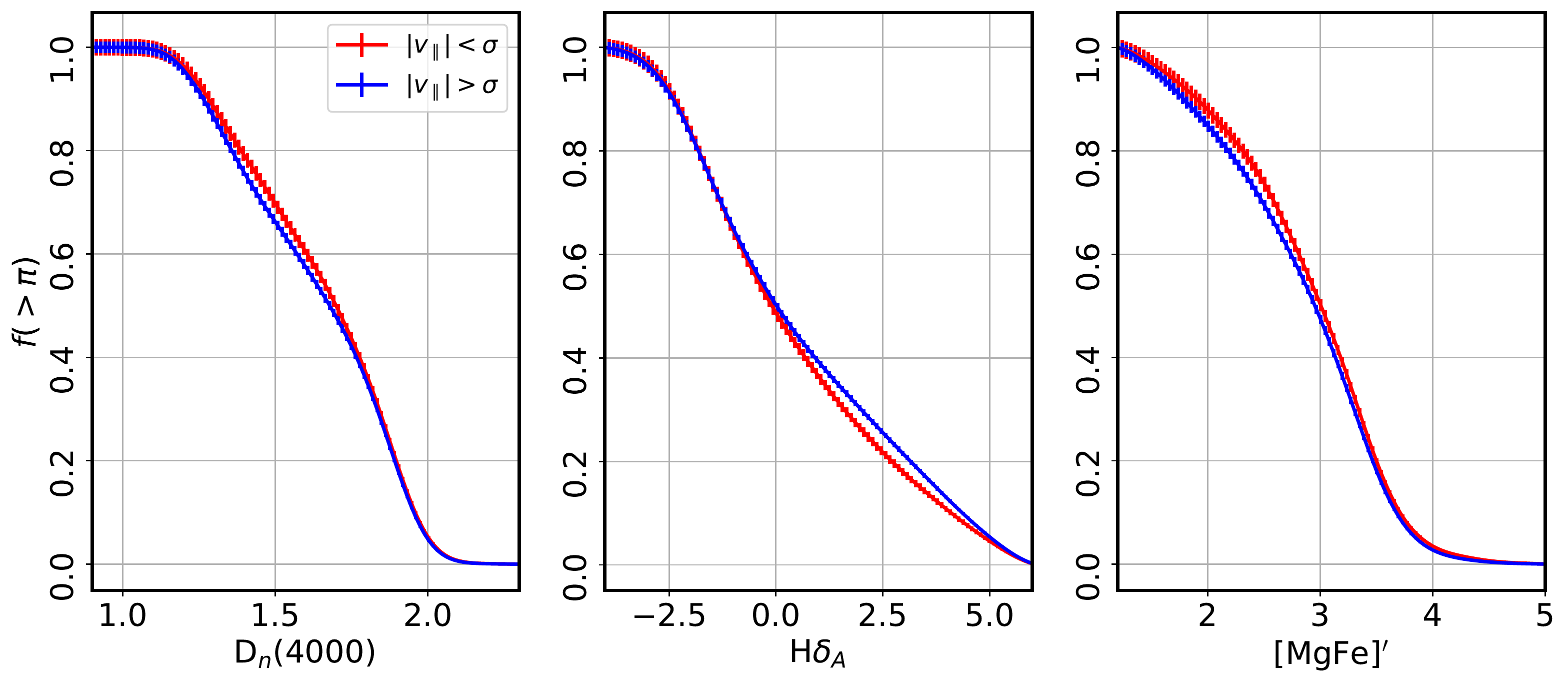}
\caption{Illustration of the cumulative functions used to define $\Delta f$
  that allows us to quantify the fraction of backsplash galaxies.
  Each panel shows the cumulative function $f(>\pi)$, where $\pi$
  represents one of the three spectral indices adopted in this paper,
  as labelled. The blue (red) lines correspond to the subsets with high- (low-)
  line-of-sight velocity. The small shaded regions are the expected Poisson
  uncertainties. Fig.~\ref{fig:histdiff} is produced
  by differences of these cumulative functions with respect to velocity,
  reflecting the contribution of backsplash galaxies (see text for details).}
  \label{fig:cumul}
\end{figure}

Our method to quantify the presence of backsplash galaxies only relies on phase space,
with the main assumption that high velocity galaxies outside of the virial radius
are infallers. In this case, the difference between high- and low-velocity
galaxies outside of the group radius will determine the fraction of backsplash
galaxies. The real scenario will of course include a number of backsplash galaxies
at high velocity, therefore our estimates will provide a robust lower bound on this
fraction. We use the distributions
presented in Fig.~\ref{fig:hist} to compute the cumulative fractions $f_v(>\pi)$, i.e.
the fraction of galaxies within a given sample that have a spectral index ($\pi$)
higher than a chosen value. This function trivially decreases from 1, at the lowest
value of $\pi$ in the sample, to 0, at the highest value.
Fig.~\ref{fig:cumul} illustrates this definition for the targeted line strengths
adopted in this paper to characterize the stellar populations of the full sample. 
To avoid crowding, we do not show the Gaussian and non Gaussian samples here.
If we define the cumulative function as $f_{v+}$ for the high velocity sample, and
$f_{v-}$ for the low velocity sample, then we expect the
difference $\Delta f\equiv f_{v-}-f_{v+}$
to represent the {\sl excess} of galaxies corresponding to the
backsplash population.  Note the differences between these cumulative
functions are non-negligible, but rather subtle.  The shaded areas in
the figure mark the standard 1\,$\sigma$ Poisson error.

\begin{figure}
\includegraphics[width=85mm]{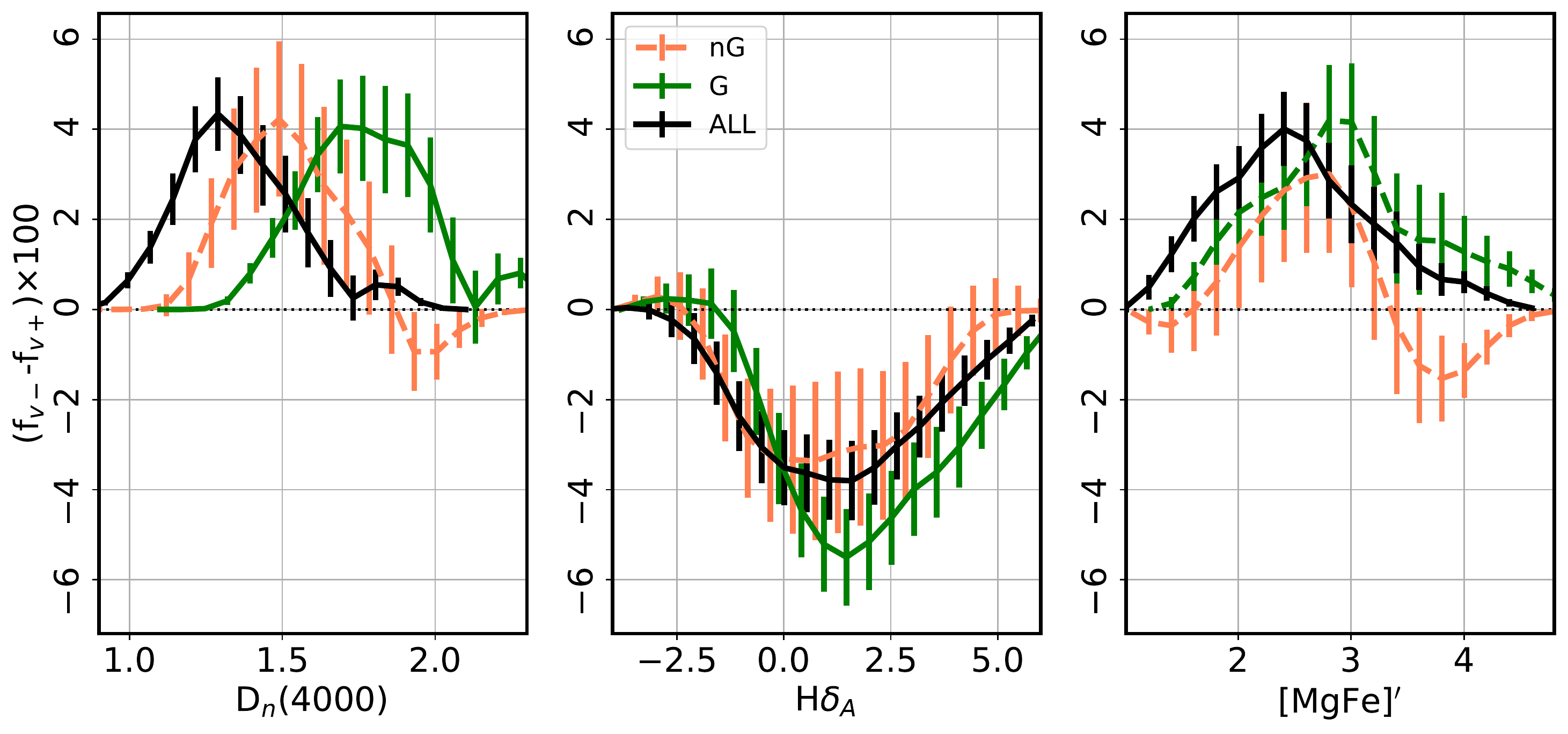}
\caption{Difference between cumulative fractions 
  of low- and high-$|v_\parallel|/\sigma$ subsets as a function
  of the three targeted spectral indices. The results are shown for
  the full cluster sample (black) as well as for Gaussian (green) and
  non Gaussian clusters (orange). See text for details.}
  \label{fig:histdiff}
\end{figure}

Fig.~\ref{fig:histdiff} shows $\Delta f$ for the distributions of the
three spectral indices in the general cluster sample (black), as well
as the subsets comprising galaxies in Gaussian (G, green) and non
Gaussian (nG, orange) clusters.  Note $\Delta f$ is expected to
trivially converge to zero at both ends, as all cumulative functions
reach 1 at low values of the spectral index, and 0 for high
values. The peak of these lines -- reminiscent of the
Kolmogorov-Smirnov (KS) statistic -- corresponds to the maximum
difference between the subsamples segregated with respect to projected
velocity, and therefore represents a lower bound in the fraction of
backsplash galaxies. While there are some variations in the
distributions between galaxies in Gaussian and non Gaussian clusters,
the difference is not statistically significant (confirmed by the $p$
values of the KS test around 0.3-0.4).

\begin{figure}
\includegraphics[width=85mm]{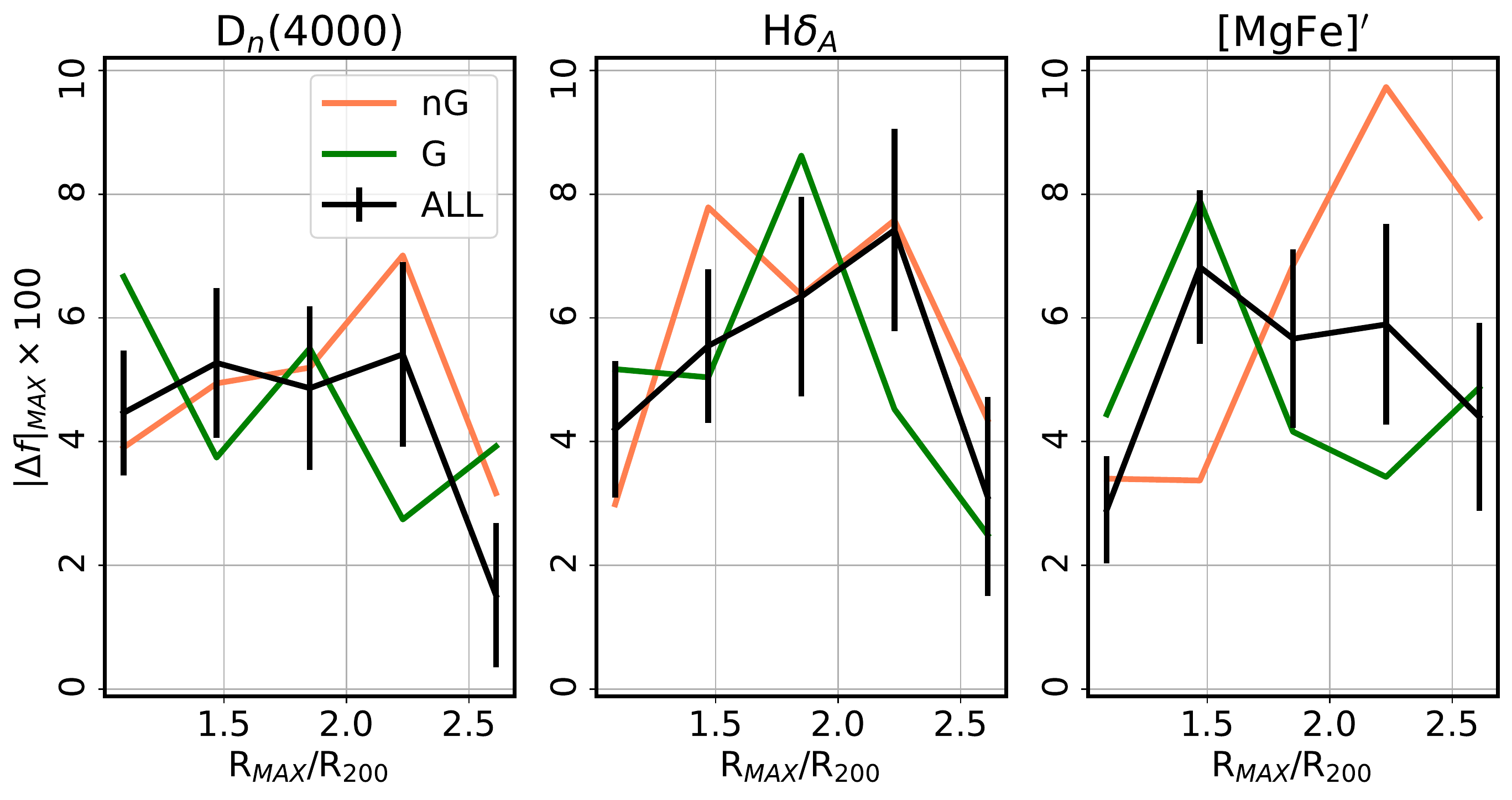}
\caption{Maximum fractional difference in line strength between the
  low- and high-velocity subsamples, a proxy for the fraction of
  backsplash galaxies, shown as a function of projected
  cluster-centric distance. The results are shown for the full sample
  (black), and for Gaussian (green) and non Gaussian clusters (orange).}
  \label{fig:fMAX}
\end{figure}

To quantify this fraction in more detail, we
show in Fig.~\ref{fig:fMAX} the dependence of this peak in the difference
between distributions ($|\Delta f|_{\rm MAX}$) and the radial interval chosen: 
Fig.~\ref{fig:histdiff} was computed for galaxies within R$\in[1,3]$R$_{\rm 200}$,
whereas in this new figure the maximum fraction is derived within a sliding 
interval $\Delta R/R_{\rm 200}=0.4$. We also include the associated Poisson
error bar for the lines corresponding to the complete set of galaxies (i.e.
not split into Gaussian and non Gaussian). Note that, although the error
bars are substantial in this detailed analysis, the difference consistently reaches a value
$|\Delta f|_{\rm MAX}\sim 0.05$ at R$\lesssim$2\,R$_{\rm 200}$. Therefore, we
claim that the fraction of backsplash galaxies {\sl with a well-defined signature}
on the stellar populations, is at least 5\% of the total population
located within 1 and 2 group radii. This figure is lower
than the predictions made by simulations: \citet{Gill:05} estimated around
half of the galaxies within 1-2$\times$R$_{\rm 200}$ to be backsplash systems, and
\citet{Haggar:20} claim that the backsplash fraction would reach $\sim$40--80\%
in low-redshift clusters, with a strong dependence on cluster-centric distance
and dynamical state. Their results are based on simulations and allow them to
precisely track the orbits of the cluster members. Therefore, the backsplash
``dynamical'' signature is determined accurately in the simulations.
Our observations would 
suggest that a dominant portion of backsplash galaxies feature similar
stellar population properties as the infallers, i.e. only a small fraction
shows signatures of stronger quenching caused by having
passed already through the cluster core. The substantial difference between
our results and those from \citet{Haggar:20} -- that can cleanly define the
set of backsplash galaxies based on their orbits -- imply that the
passage through the cluster core only produces noticeable differences
on the stellar populations in a relatively small fraction of galaxies. 
A previous set of simulations \citep{Oman:13} suggest potentially
lower fractions of backsplash galaxies. Their figure~5 shows the distribution
of infall times in different locations of PPS. If we focus on group-centric
distances larger than the virial radius, and infall times
$\gtrsim$4\,Gyr to represent the backsplash population, we only find fractions
comparable with \citet{Haggar:20} at R$\sim$R$_{200}$ but not at 1.5\,R$_{200}$.
Therefore, this type of observable constraint
can be used to calibrate galaxy formation models, both on the dynamical
properties as well as the hydrodynamics that controls the flow of the 
gas components that eventually feed star formation.
Environment-related effects on the star formation histories
of galaxies leave a signal on the stellar populations that can be tested
in samples defined in projected phase space in the same way as presented
in this paper. Phenomenological models have suggested extended quenching
timescales in agreement with our results \citep[e.g.,][]{Reeves:22}. In other words, the
observational data suggests weak (or delayed) quenching in backsplash
galaxies, in line with mechanisms based on the 'delayed-then-rapid'
environment quenching \citep{Wetzel:13}, where the slow quenching process can extend over
$\sim$2-4\,Gyr, and thus over longer than the typical dynamical timescales. Therefore,
our ansatz is only sensitive to a fraction of the total sample of backsplash
galaxies.

A more detailed view of the differences found in the line strengths
can be produced with bivariate plots of line strengths. 
We show in Fig.~\ref{fig:D4kHdA} the
distribution of cluster galaxies on the plane spanned by
[MgFe]$^\prime$ (top) or H$\delta_A$ (bottom) vs 4000\AA\ break 
strength. These diagrams are powerful indicators of the 
population properties, as D$_n$(4000)
preferentially traces the average stellar age, whereas Balmer
absorption is sensitive to recent episodes of star formation
\citep[see, e.g.][]{Kauffmann:03,Gallazzi:05}. While [MgFe]$^\prime$
is usually adopted as a metallicity indicator, it features a
significant age dependence \citep[see, e.g. fig.~4 of][]{FLB:13}.
The sample is once more 
separated into low ($|v_\parallel|<\sigma$, red) and high projected
velocity ($1\,\sigma<|v_\parallel|<3\,\sigma$, blue). The contours
represent fractions of the total number of galaxies in each sample, as
labelled. The field sample is also shown, only at the 75\% level, to
avoid crowding, with a back dashed line. 
As reference, we include as a grey line the estimates for a
set of synthethic simple stellar populations (SSP) from the models of \citet{AV:10} at
solar metallicity. The star symbols mark the values for ages of 1, 2, 3,
4, 8 and 10\,Gyr (with the youngest population represented by a white star).
The high velocity sample -- expected to be dominated by pre-processed,
infalling galaxies -- has a clear tail towards younger ages (weaker
D$_n$(4000), and higher H$\delta_A$ absorption) and lower metallicity
(although [MgFe]$^\prime$ also decreases towards young age), whereas the low
velocity sample -- which is expected to include backsplash galaxies --
is more concentrated towards weaker Balmer absorption and [MgFe]$^\prime$,
as well as stronger 4000\AA\ break strength, suggesting that this set
has more complex star formation histories, and likely having an increased
fraction of galaxies with quenched star formation,
lacking systems with SSP-equivalent ages younger
than $t_{\rm  SSP}\lesssim$2\,Gyr.  Our results align with the work of
\citet{Mahajan:11} where a special set of galaxies, termed GORES
(Galaxies with Ongoing or Recent Efficient Star formation), was
defined based on their 4000\AA\ break strength ($>$1.5) and H$\delta$
equivalent width ($>$2\AA), indicative of recent quenching.  Their
analysis, combining SDSS data with simulations, found that $\sim$19\%
of backsplash galaxies have a GORES signature, in contrast with
$\sim$34\% for the infallers, which implies quenching operates
efficiently after a core passage, but leaves open the issue that many
of the infallers may also have signatures of quenching due to
pre-processing. Our work qualitatively goes in the same direction,
but the effect is more nuanced as seen in the bivariate distribution on the
bottom-left panel of Fig.~\ref{fig:D4kHdA}. Rather than simple
thresholds on 4000\AA\ break strength and Balmer absorption, the differences,
rather affect the offset with respect to the  correlation between
H$\delta_A$ and D$_n$(4000) typically found in galaxies. The low-velocity subset
features a higher scatter. Note that simple star formation histories, defined 
by a single burst of star formation, will populate the lines traced by the stars
in the figure. In contrast, the presence of recently quenched episodes of star
formation will result in departures away from the locus described in this
diagram by the SSP prediction (see, e.g. figure~7 of \citealt{Wild:07}).
Therefore, a marked difference between the low- and high-velocity subsets
concerns galaxies with a more complex, recent history of star formation (and quenching),
as expected in a backsplash scenario.

If we take the 75\% contour level of the high
velocity subsample of the H$\delta_A$ vs D$_n$(4000) plot as reference
(thick red line) and integrate the number density {\sl outwards}, we find
a fraction of the total of 25\% for the high velocity sample (trivially,
by construction), and 58\% for the low velocity set, i.e. an excess of
33\%. The equivalent comparison in the [MgFe]$^\prime$ vs D$_n$(4000)
plot yields an excess of 19\% at low $|v_\parallel|/\sigma$, which could
be interpreted as the fractional contribution from backsplash galaxies.
We also note that this result is not dependent on the signal-to-noise
threshold imposed: samples cut at higher values produce
a consistent difference in the distributions of low- and high-velocity:
if we restrict the spectra to S/N$>$20 (measured as the median value
in the SDSS-$r$ band), the excess fractions are 27\%
(for the H$\delta_A$ vs D$_n$(4000) diagram) and 15\%
(for the [MgFe]$^\prime$ vs D$_n$(4000) distribution).

Note that a bivariate analysis is needed to produce this result,
whereas the one dimensional work presented above gives a lower 5\%
fractional difference between the high- and low-velocity
subsamples. This is corroborated by the one dimensional histograms
shown in Fig.~\ref{fig:D4kHdA}, where the differences between the two
subsets are much smaller than those in the bivariate plot. Somehow, the
2D analysis allows us to go beyond a simple interpretation of the
results as a single parameter (say the age of an SSP). A more detailed,
but model-prone, analysis is beyond the scope of this paper, but will be 
explored in a future work.

\section{Conclusions}
\label{Sec:Conc}

In this paper, we explore a method to assess the role of backsplash
galaxies in the outskirts of galaxy groups, by studying the imprint
of environment-related effects on the stellar populations. We define 
a sample of 11,252 galaxy spectra from the legacy Sloan Digital Sky Survey,
cross-matched with the groups catalogue of \citet{Yang:07}, only selecting
galaxies with cluster-centric distance 1$\lesssim$R/R$_{\rm 200}\lesssim$3, 
and separate the sample with respect to 
line of sight velocity, normalized by the 
velocity dispersion of the individual clusters. We split the
sample into low ($|v_\parallel|/\sigma<1$) and high (1$<|v_\parallel|/\sigma<$3) velocity,
and adopt the ansatz that the former includes a substantially higher fraction of backsplash
galaxies, as suggested by simulations \citep{Haggar:20}. A comparison is
also made with a general sample of SDSS galaxy spectra selected in the same
redshift window, but irrespective
of environment, always making sure the samples have the same distribution in 
stellar velocity dispersion, which is a well-known driver of population
properties in galaxies \citep{SAMIGrad:19}.

\begin{figure}
\includegraphics[width=85mm]{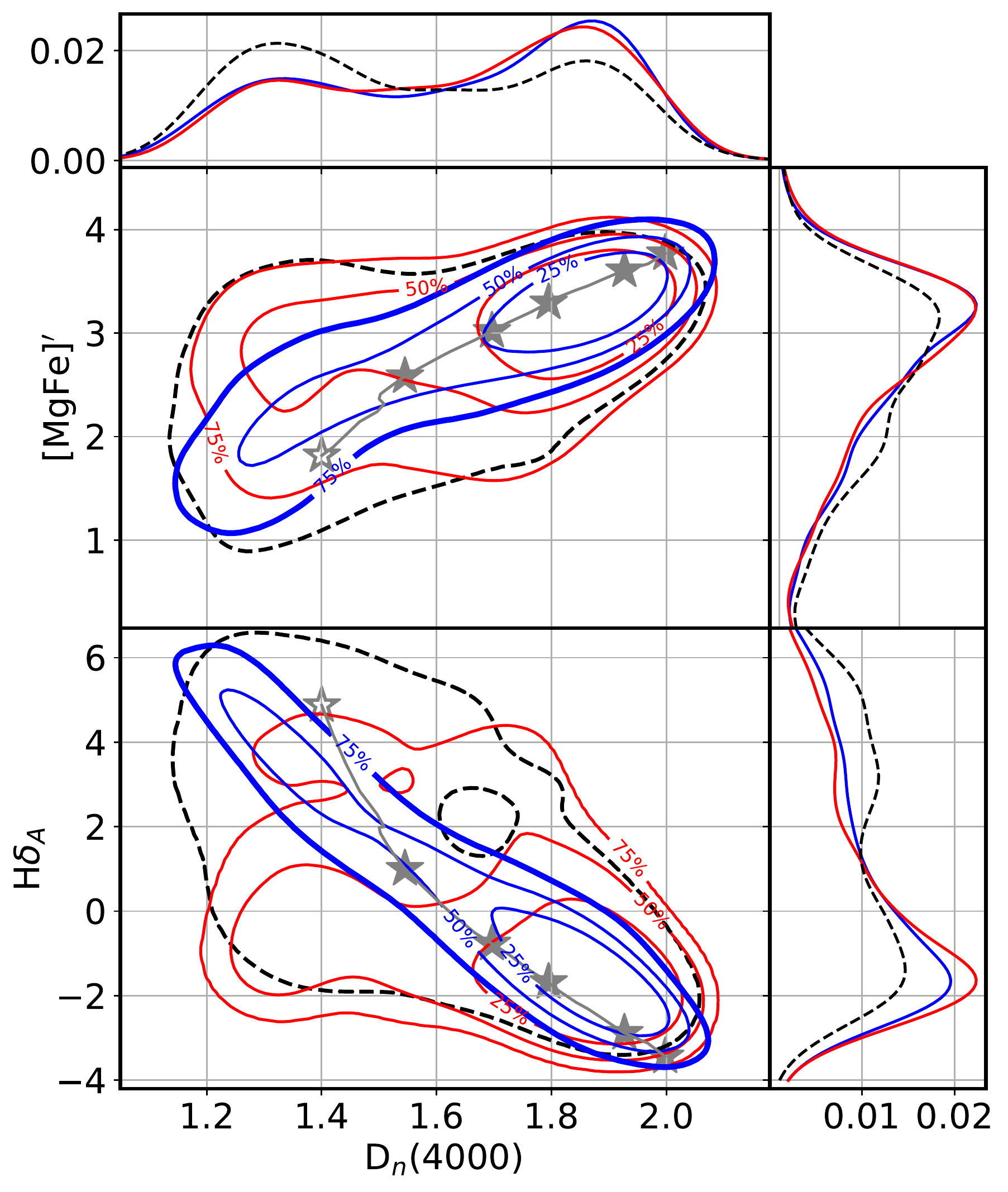}
\caption{Bivariate plots with respect to the three line strengths explored
  in this paper. The high (blue) and low (red) $|v_\parallel|/\sigma$
  subsamples are shown as a
  density plot encompassing a fraction of the total, as labelled.
  Model results for the simple stellar population models of \citet{AV:10},
  at solar metallicity, are shown, marking the ages with star symbols at 
  1, 2, 3, 4, 8, and 10\,Gyr -- with the youngest age represented by
  a white star. The 1D histograms of the three line strengths are
  also shown by the side. For reference, the black dashed lines correspond to
  the field sample -- homogenised  to the velocity dispersion distribution
  of the high $|v_\parallel|/\sigma$ sample (the contour is only shown
  at the 75\% level to avoid crowding).}
  \label{fig:D4kHdA}
\end{figure}

Our analysis confirms the trend towards older stellar ages in the
low velocity sample that we interpret as the additional contribution from the backsplash
population. This difference is consistent in three key
spectral indices: D$_n$(4000), H$\delta_A$, and [MgFe]$^\prime$,
but also manifests itself in the gas phase via lower
star formation rates and a higher fraction of quiescent galaxies in the
low velocity subsample. The difference is small but statistically significant,
which implies backsplash galaxies cannot be the dominant population, at least
regarding the observable signatures of interaction. The comparison with the
general sample confirms the very large difference between field galaxies and
those in the catchment area of clusters, confirming the strong effect of
pre-processing on the stellar content of galaxies \citep[e.g.][]{RP:17}.

By defining the difference between the cumulative fractions of galaxies with
a given spectral indicator, we quantify the role of backsplash galaxies and
conclude that the fraction of this type of galaxies should be at least 5\%. 
By comparing this fraction with respect to cluster-centric distance, we confirm
that the difference is highest around R$\lesssim$2\,R$_{\rm 200}$.
The extension of the analysis to bivariate line strength diagrams (Fig.~\ref{fig:D4kHdA}),
shows a more marked difference between the two subsamples, with the low-velocity
set featuring a wider scatter in H$\delta_A$ and higher values of
[MgFe]$^\prime$, interpreted as the contribution of backsplash galaxies,
with a more complex formation history, consistent with the quenching of
a recent episode of star formation.
Integrating the 2D number density
on these plots, we semi-quantitatively estimate the contribution of backsplash galaxies
in the range 15-30\%, more in line with numerical simulations.

Our work confirms the
presence of the backsplash population, and suggests that detailed analysis,
beyond simple 1D distributions, should be adopted to find the subtle features
left by the cluster core passage on the absorption spectra that reflects the
variations in stellar populations with respect to the infalling galaxies. 
Pre-processing is found to produce as strong a signature as in backsplash
galaxies, thus complicating observational constraints. Our estimate 
of the backsplash fraction is, at most, around $\sim$25\%, whereas
simulations suggest a higher fraction,
confirming that various quenching processes operate in the vicinity of clusters,
and backsplash quenching may be affected by delays longer than the dynamical timescale, 
hindering accurate estimates of the backsplash galaxies from spectroscopic data alone.

\section*{Acknowledgements}
IF acknowledges support from the Spanish Research Agency of the
Ministry of Science and Innovation (AEI-MICINN) under grant 
PID2019-104788GB-I00. KU acknowledges support from the Ministry of Science and Technology
of Taiwan (grant MOST 109-2112-M-001-018-MY3) and from the Academia
Sinica (grant AS-IA-107-M01). VMS acknowledges the FAPESP scholarship
programme through grants 2020/16243-3  and 2021/13683-5. 
Funding for SDSS-III has been provided by
the Alfred P. Sloan Foundation, the Participating Institutions, the
National Science Foundation, and the U.S. Department of Energy Office
of Science. The SDSS-III web site is http://www.sdss3.org/.

\section*{Data availability}
This work has been fully based on publicly available data: 
galaxy spectra were retrieved from the SDSS DR16 archive
(https://www.sdss.org/dr16/) and stellar population
synthesis models can be obtained from the respective authors.

\bibliographystyle{mnras}
\bibliography{PPSplash_Final}

\bsp	
\label{lastpage}
\end{document}